\documentclass[doublespaced,a4,12pt]{article}
\usepackage{amssymb}
\usepackage[polish, english]{babel}
\usepackage[cp1250]{inputenc}
\usepackage{amsmath,graphicx,epsf,pstricks,amsfonts,epsfig,latexsym,wasysym}
\title{\bf Vacuum pressure, dark energy and dark matter}

\author{Bogus{\l}aw Broda\footnote{bobroda@uni.lodz.pl}\; and\;
Micha{\l} Szanecki\footnote{michalszanecki@wp.pl}\\
\small\textit{Department of Theoretical Physics,}
\small\textit{University of {\L}\'od\'z}\\
\small\textit{Pomorska 149/153, PL--90--236 {\L}\'od\'z, Poland}}
\date{}
\begin{document}
\maketitle
\begin{quote}
\noindent
\begin{flushleft}
 \\
\vspace{0.2cm} \textbf{PACS:} 95.36.+x dark energy, 04.62.+v quantum fields in curved spacetime, 04.60.-m quantum gravity---in general relativity and gravitation, 98.80.Es observational cosmology, 95.35.+d dark matter.\\

\end{flushleft}
It has been argued that the correct, i.e.\ positive, sign of
quantum vacuum energy density, or more properly, negative sign of
quantum vacuum pressure, requires not a very large, and to some
extent model-independent, number, e.g.\ $\sim 100$, of additional,
undiscovered fundamental bosonic particle species, absent in the
standard model. Interpretation of the new particle species in
terms of dark matter ones permits to qualitatively, and even
quantitatively, connect all the three concepts given in the title.
\end{quote}

Dark energy \cite{Reiss} and dark matter \cite{Bertone} are two
main constituents of our Universe. Their contribution amounts to
almost $96\%$ of the total energy-mass od the Universe. The nature
of the both seems to be mysterious but completely different.
Another mysterious and elusive constituent of our Universe,
appearing in theoretical context rather than in a cosmological
one, is quantum vacuum \cite{Volovik}. But recently, mainly due to
advances in the Casimir effect, the quantum vacuum is beginning to
enter reality \cite{Jaffe}. The old idea to identify dark
energy and quantum vacuum energy is theoretically very attractive
but the main difficulty is to reconcile the values of the both
energies \cite{Weinberg},\cite{Carrol}. That is a puzzle. But
there is another puzzle, usually not being mentioned. Since, as it
is well known, the sign of quantum vacuum energy follows from the
statistics of fluctuating fields, the right (positive) sign of
vacuum energy corresponds to bosonic modes. But the number of
different fermionic particle species prevails in the standard
model and that is the puzzle. Thus, actually, we have the two
independent puzzles related to the connection between quantum
vacuum energy and dark energy: the puzzle of the huge (absolute)
value of quantum vacuum energy density, and the puzzle of its sign.

It appears, and this is the main subject of our letter, that it is
possible to solve the second puzzle, establishing a link between
the issue of dark energy and dark matter. Namely, not a very
large, and to some extent model-independent, number of
undiscovered bosonic fields should be included in the fundamental
set of particle species to obtain the right sign of vacuum energy
density. These new species are natural candidates for dark matter
particle species.

A typical approach to quantum vacuum energy yields the standard
formula \cite{Weinberg}
\begin{align}
  \varrho_{\rm vac}=\frac{1}{2}\int\limits_{0}^{\Lambda_{\rm \textsc{uv}}}
  \frac{4\pi}{(2\pi\hbar)^3c}\;
  \sqrt{(mc)^2+k^2}\;k^2\mathrm{d}k,
  \label{Vacuum energy 1}
\end{align}
where $m$ is the mass of a bosonic mode
and $k$ is its momentum. 
For a large ultraviolet
(UV) cutoff $\Lambda_{\rm \textsc{uv}}$ we have approximately
\begin{align}
  \varrho_{\rm vac}\approx\frac{1}{(4\pi)^2}\frac{{\Lambda_{\rm
  \textsc{uv}}}^4}{\hbar^2
  c}.
  \label{Vacuum energy 2}
\end{align}
Setting $\Lambda_{\rm \textsc{uv}}=\Lambda_{\rm P}$, where the
Planck momentum
\begin{align}
 \Lambda_{\rm P}=\sqrt{\frac{\hbar c^3}{G}}\approx 6.5\rm\,kg\,m/s,
 \label{Planck momentum value}
\end{align}
and $G$ is the Newton gravitational constant, we obtain the
``(in)famous'' value (formula)
\begin{align}
 \varrho_{\rm vac}\approx\frac{c^5}{(4 \pi)^2 \hbar G^2}\approx
 3.4\times10^{94}\rm\,kg/m^3
\approx\frac{ M_{\rm P}}{L_{\rm P}^3}.
 \label{Vacuum energy 3}
\end{align}
where $M_{\rm P}$ and $L_{\rm P}$ is the Planck mass and length, respectively.
The fame of this formula rivals its absurdity. Not only is
\eqref{Vacuum energy 3} some $10^{120}$ times greater than
expected but evidently the sign is not as expected. For
fermionic modes the sign of \eqref{Vacuum energy 3} will be
reversed! Therefore, for presently known contents of fundamental
set of fields (with prevalent fermionic modes) the sign will be
wrong. Consequently, new bosonic species are urgently being looked
for.

One should emphasize that the difficulty with the sign seems to be
independent of the approach to the issue of the huge value of
quantum vacuum energy. In other words, additional bosonic modes
are presumably unavoidable provided quantum vacuum is supposed to
have something to do with dark energy or, at least, with reality.
Taking for granted that new bosonic particle species should enter
the set of fundamental fields any further estimate of their number
could already depend on the assumed model of dark energy in the
framework of the idea of quantum vacuum.

In \cite{Broda}, we have proposed a phenomenologically
promising approach to solve the puzzle of the huge (absolute)
value of ``the quantum vacuum energy density''. In the framework
of our approach the (absolute) value of the ``vacuum energy
density'' of a single mode is of the order
\begin{align}
\varrho\sim 0.01\varrho_{\rm exp}, \label{Vacuum energy density}
\end{align}
where $\varrho_{\rm exp}$ is the experimental value of the energy
density of dark energy. Thus, Eq.~\eqref{Vacuum energy density}
gives the result for a single mode. Obviously, the relation
\eqref{Vacuum energy density} is model dependent. In our approach
\cite{Broda} it is just \eqref{Vacuum energy density}.

Strictly speaking, the formula yielding the result \eqref{Vacuum
energy density}, i.e.\ the lagrangian density \cite{Broda}
\begin{align}
\mp\frac{1}{4}\;\frac{1}{(4\pi)^2 G}\;\frac{1}{2}(1-q){H}^2,
\label{Rho formula}
\end{align}
where
the upper and lower sign corresponds to a bosonic and fermionic mode, respectively,
$q$ is the present deceleration parameter, and $H$ is the
present Hubble expansion rate, corresponds to the pressure rather
than to the energy density.

In fact, the diagonal part of the energy-momentum tensor,
\begin{align}
T_{\mu\nu}=\frac{\partial\mathcal{L}}{\partial\;\partial^{\mu}\phi}\partial_{\nu}\phi-g_{\mu\nu}\mathcal{L}
\label{Energy-momentum tensor}
\end{align}
reduces to:
\begin{align}
 T_{00}\equiv \varrho=-g_{00}\mathcal{L},\;\; \rm{for} \;\;
\partial_{0}\phi=0;\label{Tensor diagonal terms a}
\end{align}
or
\begin{align}
 T_{ii}\equiv p=-g_{ii}\mathcal{L},\;\; \rm{for}\;\;
\partial_{i}\phi=0 \label{Tensor diagonal terms b},
\end{align}
where $c=1$ and the signature of the metric is
$\left(+,-,-,-\right)$. Our case corresponds obviously to the
second possibility, i.e.~\eqref{Tensor diagonal terms b}, because
our fields are homogeneous $(\partial_{i}\phi=0)$ but
time-dependent (Friedmann--Lema\^{\i}tre--Robertson--Walker
cosmological model).

Coming back to our model
\begin{align}
\mp\frac{1}{4}\;\frac{1}{(4\pi)^2
G}\;\frac{1}{2}(1-q){H}^2\approx\mathcal{L}=-\frac{1}{g_{ii}}\;p,
\label{Lagrange correspondence with pressure}
\end{align}
and therefore
\begin{align}
p\approx\mp\frac{1}{4}\;\frac{1}{(4\pi)^2
G}\;\frac{1}{2}(1-q){H}^2, \label{Pressure/density formula}
\end{align}
which still conforms with our description of dark energy in terms
of quantum vacuum. For example, assuming temporarily an ad hoc
barotropic relation
\begin{align}
p=w\varrho,\;w=-1,
\label{Barotropic relation}
\end{align}
we simply obtain
\begin{align}
\varrho\approx\pm\frac{1}{128{\pi}^{2} G}(1-q){H}^2.
\label{Density from barotropic formula}
\end{align}
But Eq.~\eqref{Pressure/density formula} (for pressure) is more
fundamental than Eq.~\eqref{Density from barotropic formula} (for
energy density) because no equation of state need to be presumed.

Consequently, as a next step, we should collect contributions, of
the type estimated by us, coming from all fundamental physical
modes. We can proceed in the spirit of the philosophy of quantum
(vacuum) induced interactions (see,
e.g.~\cite{Sakharov},\cite{Broda3}). Thus, the total pressure,
coming from all fundamental modes is according to
\eqref{Pressure/density formula} of the order
\begin{align}
p\approx-\frac{N_{0}}{128{\pi}^{2} G}(1-q){H}^2, \label{Pressure
with modes}
\end{align}
and
$\varrho_{\rm exp}\approx N_{0}\varrho$,
where $N_{0}$ is the ``alternated sum'' of the fundamental modes.
Namely, we define
\begin{align}
N_{0}\equiv n_{B}-n_{F}=\sum_{k=0}(-)^{k}n_{k/2}, \label{Modes
number definition}
\end{align}
where $n_{B}$ and $n_{F}$ is the number of fundamental bosonic and
fermionic modes, respectively. By virtue of the spin-statistics
theorem we can rewrite $N_{0}$ in terms of spin degrees of
freedom, i.e.\ $n_{0}$, $n_{1/2}$, $n_{1}$, etc.
One should note that the particular case
$n_{B}=n_{F}$
corresponds to supersymmetry.

Recapitulating, the puzzle related to quantum vacuum fluctuations
consists in getting under control the huge value of the vacuum
energy density, strictly speaking, from our point of view, the (absolute)
value of the pressure. Once the value becomes reasonable in its
size, the sign puzzle emerges. The sign assumes the expected
value, e.g.\ ``minus'' in \eqref{Pressure with modes}, only
provided the number of bosonic fundamental modes is prevalent.

We would like to stress, once more, that the conclusion concerning
bosonic species is not specific to our model of dark energy,
because only bosonic modes give contributions huge or moderate but
with right signs. Since, the standard model contains greater
number of fermionic particle species (leptons and quarks) than
bosonic ones (mainly, gauge fields), we can conclude that there is
a missing number of invisible bosonic modes. The bosonic modes do
not enter the standard model but they must interact
gravitationally. Therefore, they are appropriate and natural
candidates to the role of dark matter particles.

More concretely, for the standard model
\begin{align}
N_{0}=[4]-2\cdot3\cdot\left(3+3\cdot4)+2\cdot(1+3+8+[1]\right)=-(63\pm3),
\label{Modes number exact calculus}
\end{align}
where in the first bracket we have included, as yet not
discovered, Higgs modes, next there are leptons and quarks
(2 spins $\times$ 3 families $\times$ (3 leptons of the both helicities $+$ 3 colors $\times$ 4 quarks of the both helicities)),
and
finally, gauge fields
(2 spins $\times$ (photon $+$ 3 weak bosons $+$ 8 gluons))
with graviton in the last bracket. From
\eqref{Pressure with modes} and \eqref{Modes number exact
calculus} (see, also Eq.~\eqref{Vacuum energy density} for a
numeric value) it follows that the lacking number of bosonic modes
is of the order $n_{B}\sim100$. This number is model-dependent but
the conclusion is not.

Any possible, alternative and independent solution of the puzzle
of the huge (absolute) value of quantum vacuum energy density also
would require additional bosonic species but their number could
vary. For example, any potential solution directly yielding the
right (absolute) value of the quantum vacuum energy density,
meaning that the correct (absolute) value is recovered for a
single mode, would require a tuning between the number bosonic and
fermionic species, i.e.\ $n_{B}-n_{F}\sim 1$.

Certainly, it would be very advantageous to present an alternative
and independent estimate of the number of the lacking fundamental
bosonic modes $n_{B}$. To this end we will make use of the
proposal given in \cite{Broda3}, concerning quantum vacuum induced
gravitational action and quantum vacuum induced gravitational
(black hole) thermodynamical entropy. Assuming the number and kind
of fundamental modes given in \eqref{Modes number exact calculus},
we get $n_{B}\sim 40$ to properly induce the gravitational action
(see Eq.~(8) in \cite{Broda3}), and $n_{B}\sim 200$ to properly
induce the gravitational (black-hole) thermodynamical entropy (see
Eq.~(2) in \cite{Broda3}). Taking into account an approximate
character of our reasoning one should admit that it qualitatively
agrees with our present dark energy estimate, i.e.\
$n_{B}\sim100$.

In this letter, we have proposed a consistent connection between
the following three concepts: quantum vacuum pressure, dark energy
and dark matter. First of all, we have shown that, in the context
of lagrangian approach, the quantum pressure naturally replaces
the notion of quantum vacuum energy density. Next, we
have argued that, independently of an actual model of dark energy,
additional, undiscovered fundamental bosonic particle species are
necessary. Finally, applying our earlier model of dark energy
\cite{Broda} we have estimated the number
of the fundamental bosonic particle species, namely
$n_{B}\sim100$. Amazingly, it agrees with the estimates being yielded by
induced gravity \cite{Broda3}.

\section*{Acknowledgments}
This work has been supported by the University of {\L}\'od\'z
grant and LFPPI network.

\end{document}